\documentclass[aps,prl,showpacs,showkeys,twocolumn,
 superscriptaddress,altaffilletter,amsmath,amssymb]{revtex4-1}

\usepackage{graphicx}
\usepackage{upgreek}
\usepackage{xcolor}
\usepackage[mathlines]{lineno}

\begin{document}

\title{Final Results of GERDA on the Search for Neutrinoless Double-$\beta$ Decay}

\collaboration{{{\sc Gerda} Collaboration}}
\email{correspondence: gerda-eb@mpi-hd.mpg.de}
\noaffiliation

  \affiliation{INFN Laboratori Nazionali del Gran Sasso and Gran Sasso Science Institute, 67100 Assergi, Italy}
  \affiliation{INFN Laboratori Nazionali del Gran Sasso and Universit{\`a} degli Studi dell'Aquila, 67100 L'Aquila,  Italy}
  \affiliation{INFN Laboratori Nazionali del Sud, 95123 Catania, Italy}
  \affiliation{Institute of Physics, Jagiellonian University, 31-007 Cracow, Poland}
  \affiliation{Institut f{\"u}r Kern- und Teilchenphysik, Technische Universit{\"a}t Dresden, 01069 Dresden, Germany}
  \affiliation{Joint Institute for Nuclear Research, 141980 Dubna, Russia}
  \affiliation{European Commission, JRC-Geel, 2442 Geel, Belgium}
  \affiliation{Max-Planck-Institut f{\"u}r Kernphysik, 69117 Heidelberg, Germany}
  \affiliation{Department of Physics and Astronomy, University College London, London WC1E 6BT, United Kingdom}
  \affiliation{Dipartimento di Fisica, Universit{\`a} Milano Bicocca, 20126 Milan, Italy}
  \affiliation{INFN Milano Bicocca, 20126 Milan, Italy}
  \affiliation{Dipartimento di Fisica, Universit{\`a} degli Studi di Milano and INFN Milano, 20133 Milan, Italy}
  \affiliation{Institute for Nuclear Research of the Russian Academy of Sciences, 117312 Moscow, Russia}
  \affiliation{Institute for Theoretical and Experimental Physics, NRC ``Kurchatov Institute'', 117259 Moscow, Russia}
  \affiliation{National Research Centre ``Kurchatov Institute'', 123182 Moscow, Russia}
  \affiliation{Max-Planck-Institut f{\"ur} Physik, 80805 Munich, Germany}
  \affiliation{Physik Department, Technische  Universit{\"a}t M{\"u}nchen, 85748 Munich, Germany}
  \affiliation{Dipartimento di Fisica e Astronomia, Universit{\`a} degli Studi di Padova, 35131 Padua, Italy}
  \affiliation{INFN  Padova, 35131 Padua, Italy}
  \affiliation{Physikalisches Institut, Eberhard Karls Universit{\"a}t T{\"u}bingen, 72076 T{\"u}bingen, Germany}
  \affiliation{Physik-Institut, Universit{\"a}t Z{\"u}rich, 8057 Z{u}rich, Switzerland}

\author{M.~Agostini}
  \affiliation{Department of Physics and Astronomy, University College London, London WC1E 6BT, United Kingdom}
  \affiliation{Physik Department, Technische  Universit{\"a}t M{\"u}nchen, 85748 Munich, Germany}
\author{G.R.~Araujo}
  \affiliation{Physik-Institut, Universit{\"a}t Z{\"u}rich, 8057 Z{u}rich, Switzerland}
\author{A.M.~Bakalyarov}
  \affiliation{National Research Centre ``Kurchatov Institute'', 123182 Moscow, Russia}
\author{M.~Balata}
  \affiliation{INFN Laboratori Nazionali del Gran Sasso and Gran Sasso Science Institute, 67100 Assergi, Italy}
\author{I.~Barabanov}
  \affiliation{Institute for Nuclear Research of the Russian Academy of Sciences, 117312 Moscow, Russia}
\author{L.~Baudis}
  \affiliation{Physik-Institut, Universit{\"a}t Z{\"u}rich, 8057 Z{u}rich, Switzerland}
\author{C.~Bauer}
  \affiliation{Max-Planck-Institut f{\"u}r Kernphysik, 69117 Heidelberg, Germany}
\author{E.~Bellotti}
  \affiliation{Dipartimento di Fisica, Universit{\`a} Milano Bicocca, 20126 Milan, Italy}
  \affiliation{INFN Milano Bicocca, 20126 Milan, Italy}
\author{S.~Belogurov}
  \altaffiliation[also at: ]{NRNU MEPhI, Moscow, Russia}
  \affiliation{Institute for Theoretical and Experimental Physics, NRC ``Kurchatov Institute'', 117259 Moscow, Russia}
  \affiliation{Institute for Nuclear Research of the Russian Academy of Sciences, 117312 Moscow, Russia}
\author{A.~Bettini}
  \affiliation{Dipartimento di Fisica e Astronomia, Universit{\`a} degli Studi di Padova, 35131 Padua, Italy}
  \affiliation{INFN  Padova, 35131 Padua, Italy}
% ...

\date{\today}

\begin{abstract}
The GERmanium Detector Array (GERDA) experiment searched for the lepton-number-violating neutrinoless double-$\beta$ ($0\nu\beta\beta$) decay of $^{76}$Ge, whose discovery would have far-reaching implications in cosmology and particle physics.
By operating bare germanium diodes, enriched in $^{76}$Ge, in an active liquid argon shield, GERDA achieved an unprecedently low background index of $5.2\times10^{-4}$~counts/(keV\,kg\,yr) in the signal region and met the design goal to collect an exposure of 100~kg\,yr in a background-free regime.
When combined with the result of Phase~I, no signal is observed after 127.2~kg\,yr of total exposure. A limit on the half-life of $0\nu\beta\beta$ decay in $^{76}$Ge is set at $T_{1/2}>1.8\times10^{26}$~yr at 90\%~C.L., which coincides with the sensitivity assuming no signal.
\end{abstract}

\maketitle

\nolinenumbers

The matter-antimatter asymmetry of the Universe remains an important unsolved puzzle of cosmology and particle physics.
Many theories predict that the asymmetry is produced by a violation of lepton number via leptogenesis~\cite{Davidson:2008bu}.
These theories naturally lead to neutrinos being their own anti-particles and developing a Majorana
mass component.  Neutrino Majorana masses and lepton-number violation can be verified at the same time by observing a
hypothetical nuclear transition $(A,Z) \to (A,Z+2) + 2e^-$, called neutrinoless double-$\beta$ ($0\nu\beta\beta$) decay~\cite{Schechter:1981bd}. In
$0\nu\beta\beta$ decay, two neutrons in the parent nucleus convert into two protons and two electrons. 
Unlike the known neutrino-accompanied double-$\beta$ ($2\nu \beta \beta$) decay, the two
electrons emitted in a $0\nu\beta\beta$ decay would share the entire energy released in
the process. The main experimental signature of $0\nu\beta\beta$ decay is hence a
characteristic peak in the energy distribution, located at the  $Q$-value of the
decay ($Q_{\beta\beta}$).
A vigorous experimental program is underway to search for this transition  
in various candidate isotopes: 
$^{76}$Ge~\cite{Agostini:2019hzm, Alvis:2019sil}, $^{82}$Se~\cite{Azzolini:2019tta},
$^{100}$Mo~\cite{Alenkov:2019jis, Arnold:2015wpy, Armengaud:2019loe},
$^{130}$Te~\cite{Adams:2019jhp, Andringa:2015tza},
$^{136}$Xe~\cite{Anton:2019wmi, KamLAND-Zen:2016pfg, Martin-Albo:2015rhw}, and others.

In this paper, the final results of the GERmanium Detector Array (GERDA) experiment on the
search for the $0\nu\beta\beta$ decay of $^{76}$Ge are presented.
GERDA used high-purity germanium detectors made out of material isotopically enriched in
$^{76}$Ge to $\sim$87\%~\cite{Ackermann:2012xja,Agostini:2017hit}:
this approach maximizes the detection efficiency as source and detector coincide.
The outstanding energy resolution of germanium detectors guarantees a very
clear signature of the $0\nu\beta\beta$ decay signal. Background 
around $Q_{\beta\beta} = 2039.06$~keV~\cite{Mount:2010zz} was minimized by operating the bare detectors in liquid argon (LAr),
which provides both shielding and cooling~\cite{Heusser:1995wd}.

Phase~I of GERDA collected 23.5~kg\,yr of exposure (= total germanium mass $\times$ live time)
between November 2011 and September 2013, with an average
background index $B$ of $11\times10^{-3}$~counts/(keV\,kg\,yr) at
$Q_{\beta\beta}$~\cite{Agostini:2013mzu}.
Phase~II of GERDA started in December 2015, after a major upgrade~\cite{Agostini:2017hit}
with additional germanium detectors of superior performance and a LAr veto system~\cite{Agostini:2015boa}.
The goal was to reduce the background below $B=10^{-3}$~counts/(keV\,kg\,yr) 
and to collect 100~kg\,yr of exposure in a background-free regime. In this regime the most probable 
number of background events in the signal region is zero and the sensitivity scales linearly with the exposure, instead 
of the square-root.
Initially, 20~kg of broad energy germanium (BEGe) detectors~\cite{Agostini:2014hra, Agostini:2019mwn} were added to 
15.6~kg of coaxial detectors already operated in Phase\,I. 
After the last data release in 2018~\cite{Agostini:2019hzm}, additional inverted coaxial (IC)
detectors~\cite{Cooper:201125} with a total mass of 9.6~kg were installed, as summarized in
Tab.~\ref{tab:datasets}.

The GERDA experiment is located at the Laboratori Nazionali del Gran Sasso (LNGS) of INFN, Italy, where a
rock overburden of 3500~m water equivalent reduces the flux from cosmic muons by six orders of magnitude.
The array of germanium detectors is lowered in a cryostat containing 64~m$^3$ of LAr
through a lock system inside a clean room.
The cryostat is surrounded by a water tank (590~m$^3$ purified water) equipped with photomultipliers (PMTs) to detect the residual cosmic muons reaching the experiment. 
The water and LAr shield the core of the setup from external natural radioactivity and neutrons.
The muon veto system~\cite{Freund:2016fhz} is complemented by scintillator panels installed on the top of the clean room. 

The 41 germanium detectors are assembled into seven strings and each string is placed inside a nylon cylinder to
limit the LAr volume from which
radioactive ions can be collected by electric fields. 
This strategy effectively reduces the background due to the $\beta$ decay of $^{42}$K, which 
is produced as a progeny of the long-lived $^{42}$Ar and
has a $Q$-value above $Q_{\beta\beta}$~\cite{Lubashevskiy:2017lmf}.

A cylindrical volume around the array is instrumented with photosensors, which detect the scintillation light
in the LAr.
The LAr veto system consists of a curtain of wavelength-shifting fibers connected to silicon photomultipliers and 16 cryogenic PMTs~\cite{JanicskoCsathy:2010bh,Agostini:2017hit}.
During the upgrade, the geometrical coverage of the fiber curtain was improved. 

The germanium detectors are connected to charge-sensitive amplifiers located inside the LAr about 35~cm above the array.
The signals are digitized at 25~MHz for a total length of $160~\upmu$s and at 100~MHz in a 10-$\upmu$s window around the rising edge and are stored on disk for
analysis. 

The offline analysis of the digitized signals follows the procedures described in Ref.~\cite{Agostini:2011mh}.
Since Phase\,I, the GERDA Collaboration adopted a strict blinded a\-na\-ly\-sis: events with a reconstructed energy within $\pm25$~keV of $Q_{\beta\beta}$ are removed from the data stream and not analyzed further until all analysis procedures and parameters have been finalized.
The energy of the events in the germanium detectors is reconstructed with a zero-area cusp filter~\cite{Agostini:2015pta}, whose parameters are optimized for each detector and calibration run.
Weekly calibration runs with $^{228}$Th sources are performed to determine the energy scale and resolution, as well as to define and monitor the analysis cuts.
The energy resolutions, defined as full width at half maximum (FWHM), at $Q_{\beta\beta}$ of each detector type are 
summarized in Tab.~\ref{tab:datasets}, together with their standard deviations. 
The new IC detectors show an average resolution of 2.9~keV,
a remarkable achievement given their mass of $\sim$2~kg, comparable to the coaxial detectors;
in addition, they provide a similarly efficient identification of
the event topology, and hence background rejection~\cite{Domula:2017mei}, as the much smaller ($\sim$0.7~kg) BEGe detectors.
The energy resolution is stable within $0.1$~keV for most of the detectors over the full data taking period.
Gain stability and noise are monitored by test pulses injected into the front-end electronics at a rate of 0.05~Hz. 
The fraction of data corresponding to stable operating conditions that are used for physics analysis is about 80\% of the total.
Signals originating from electrical discharges or bursts of noise are rejected by quality cuts based on the flatness of the baseline, polarity and time structure of the pulse.
Physical events at $Q_{\beta\beta}$ are accepted with an efficiency larger than 99.9\%.

The two electrons emitted in a double-$\beta$~decay have a range in germanium of the order of 1~mm: they deposit their
energy in a small volume of the detector and thus produce highly localized events (single-site events, SSEs). In contrast, $\gamma$ rays
of similar energy mostly interact via Compton scattering
and can produce events with several separated energy depositions (multiple-site events, MSEs).
Events in which more than one germanium detector is fired are therefore identified as background.
The unique feature in Phase~II of GERDA is the LAr veto, that allows to reject events in which energy is deposited in the LAr volume surrounding the germanium detectors.
If any of the photosensors detects a signal of at least one photoelectron within about $6 \upmu$s of the germanium detector trigger, the event is classified as background.
Accidental coincidences lead to a dead time of $(2.3\pm0.1)\%$ ($(1.8\pm0.1)\%$) before (after) the upgrade, measured by randomly triggered events.
Events are discarded also if preceded by a muon-veto signal within $10~\upmu$s; the induced dead time is $< 0.01\%$.

\begin{table*}
    \centering
    \caption{Summary of the GERDA Phase~II parameters for different detector types and before/after the upgrade.
      The components of the total efficiency $\varepsilon$ for $0\nu\beta\beta$
      decays are reported individually. The efficiencies of muon veto and quality cuts are above $99.9\%$ and
      are not shown. Energy resolutions and all $0\nu\beta\beta$ decay efficiencies are reported as exposure-weighted
      averages for each detector type and their uncertainties are given as standard deviations.} \label{tab:datasets}
    \begin{tabular}{l|cc|ccc}
         \hline
         & \multicolumn{2}{c|}{Dec 2015 -- May 2018} & \multicolumn{3}{|c}{July 2018 -- Nov 2019} \\
         & coaxial & BEGe & coaxial & BEGe & inverted coaxial \\
         \hline
         Number of detectors & 7 & 30 & 6 & 30 & 5 \\
         Total mass & 15.6~kg  & 20~kg & 14.6~kg & 20~kg & 9.6~kg \\
         Exposure $\mathcal{E}$ & 28.6~kg\,yr & 31.5~kg\,yr & 13.2~kg\,yr & 21.9~kg\,yr & 8.5~kg\,yr \\
         Energy resolution at $Q_{\beta\beta}$ (FWHM) & $(3.6\pm0.2)$~keV & $(2.9\pm0.3)$~keV & $(4.9\pm1.4)$~keV & $(2.6\pm0.2)$~keV & $(2.9\pm0.1)$~keV \\
         $0\nu\beta\beta$ decay detection efficiency $\varepsilon$:& $(46.2\pm5.2)\%$ & $(60.5\pm3.3)\%$ & $(47.2\pm5.1)\%$ & $(61.1\pm3.9)\%$ & $(66.0\pm1.8)\%$ \\
         ~~ Electron Containment & $(91.4\pm1.9)\%$ & $(89.7\pm0.5)\%$ & $(92.0\pm0.3)\%$ & $(89.3\pm0.6)\%$ & $(91.8\pm0.5)\%$ \\
         ~~ $^{76}$Ge enrichment & $(86.6\pm2.1)\%$ & $(88.0\pm1.3)\%$ & $(86.8\pm2.1)\%$ & $(88.0\pm1.3)\%$ & $(87.8\pm0.4)\%$ \\
         ~~ Active volume & $(86.1\pm5.8)\%$ & $(88.7\pm2.2)\%$ & $(87.1\pm5.8)\%$ & $(88.7\pm2.1)\%$ & $(92.7\pm1.2)\%$ \\
         ~~ Liquid Argon veto & \multicolumn{2}{c|}{$(97.7\pm0.1)\%$} & \multicolumn{3}{|c}{$(98.2\pm0.1)\%$} \\
         ~~ Pulse shape discrimination  & $(69.1\pm5.6)\%$ & $(88.2\pm3.4)\%$ & $(68.8\pm4.1)\%$ & $(89.0\pm4.1)\%$ & $(90.0\pm1.8)\%$ \\
         \hline
    \end{tabular}
\end{table*}

The pulse shape of the germanium detector signals is used to discriminate background events.
In addition to $\gamma$-induced MSEs, events due to $\alpha$ or $\beta$ decays on the detector surface can also be identified.
In the case of the BEGe and IC detectors one pa\-ra\-me\-ter, $A/E$, is used to classify background events, where $A$ is the maximum current amplitude and $E$ is the energy. As MSEs and surface events at the n$^+$ electrode are characterized by wider current pulses, they feature a lower $A/E$ value
compared to SSEs, while surface events at the very thin ($<1 \upmu$m) p$^+$ electrode show a higher $A/E$ value~\cite{Budjas:2009zu}.
Therefore, rejecting events on both sides of the $A/E$ distribution of SSEs enhances the signal to background ratio.
The coaxial detectors feature a more complicated time structure which requires an artificial neural network (ANN) to discriminate SSEs from MSEs and a dedicated cut on the signal rise time to discard events on the p$^+$ electrode~\cite{Agostini:2013jta,Agostini:2019hzm}.

An additional cut is applied to all detectors to remove events with slow or incomplete charge collection. These events are not necessarily due to background but rather to energy depositions in particular parts of the detectors
featuring unusual charge collection dynamics. These events are identified through the difference between two energy estimates performed using the same digital filter but different shaping times.
An event is discarded if the energy difference is larger than three standard deviations from the average.

$^{228}$Th calibration data are used to train the ANN and to tune the $A/E$ discrimination.
The double escape peak (DEP) at 1593~keV of the prominent $\gamma$ ray of $^{208}$Tl at 2615~keV is used as a sample of SSEs,
and the full energy peak at 1621~keV from $^{212}$Bi as a sample of MSEs.
The MSE cut threshold is set for all detectors at $90\%$ DEP survival fraction.
The threshold to reject p$^+$ surface events is optimized using the $2\nu\beta\beta$ and $\alpha$ decays.
The $0\nu\beta\beta$ decay signal efficiency is estimated for all detectors from the survival fraction of DEP and $2\nu\beta\beta$ decay events after all cuts. An extrapolation to $Q_{\beta\beta}$ is performed to take into account the energy dependence.
The combined signal efficiency of pulse shape discrimination is reported in Tab.~\ref{tab:datasets} for each 
detector type, before and after the upgrade.

\begin{figure*}
    \includegraphics[width=2.0\columnwidth]{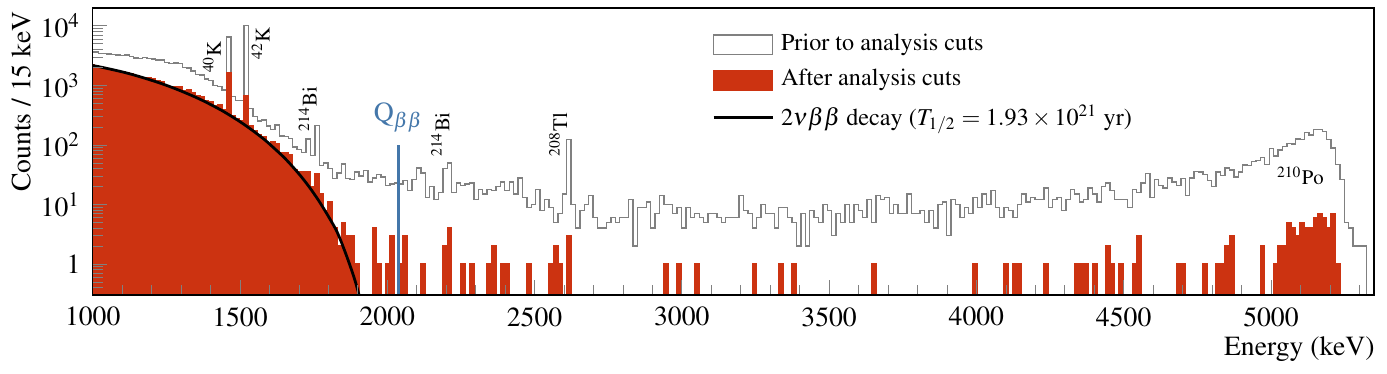}
    \caption{Energy distribution of GERDA Phase~II events (exposure of 103.7~kg\,yr) before and after analysis cuts. The expected distribution of $2\nu\beta\beta$ decay events is shown assuming the half-life measured by GERDA~\cite{Agostini:2015nwa}. The prominent $\gamma$ lines and the $\alpha$ population
    around 5.3~MeV are also labeled.}
    \label{fig:spectrum}
\end{figure*}

GERDA Phase~II data were collected between December 2015 and November 2019. The total exposure is
103.7 kg\,yr (58.9~kg\,yr already published in~\cite{Agostini:2019hzm} and 44.8~kg\,yr of new data).
Fig.~\ref{fig:spectrum} shows the energy distribution of all events before and after applying the analysis cuts.
At low energy, the counting rate is mostly accounted for by the $2\nu\beta\beta$ decay of $^{76}$Ge with a half-life of
$T_{1/2}^{2\nu \beta \beta} = (1.926 \pm 0.094) \times 10^{21}$~yr~\cite{Agostini:2015nwa}.

\begin{figure}
    \includegraphics[width=\columnwidth]{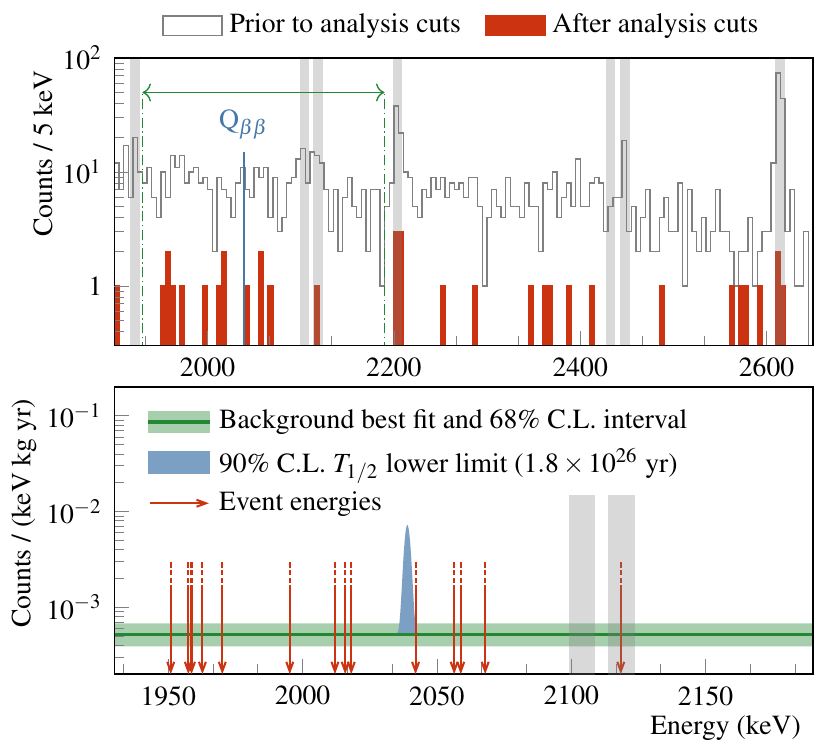}
    \caption{{\it Top:} Energy distribution of GERDA Phase\,II events before and after analysis cuts. The grey areas indicate regions in which $\gamma$ lines are expected. The dashed lines mark the edges of the analysis window. 
      {\it Bottom:} Energy of the events in the analysis window after analysis cuts.
      The blue peak displays the expected $0\nu\beta\beta$ decay signal for $T_{1/2}$ equal to the lower limit,
      $1.8 \times 10^{26}$~yr. Its width is the resolution $\sigma_k$ of the partition which contains the event closest
      to $Q_{\beta\beta}$.} \label{fig:roi}
\end{figure}

The energy range considered for the $0\nu\beta\beta$ decay a\-na\-ly\-sis goes from 1930~keV to 2190~keV, with the exclusion of the intervals $(2104\pm5)$~keV and $(2119\pm5)$~keV that contain two known background peaks (Fig.~\ref{fig:roi}).
No other $\gamma$ line or structure is expected in this analysis window according to the
background model~\cite{Abramov:2019hhh}.
After unblinding, 13~events are found in this analysis window after all cuts (5 in coaxial, 7 in BEGe and 1 in IC detectors).
These events are likely due to $\alpha$ decays, $^{42}$K $\beta$ decays, or $\gamma$ decays from $^{238}$U and $^{232}$Th series.
Data presented in~\cite{Agostini:2017iyd}, when less effective discrimination techniques against surface events in coaxial detectors were available, have been re-analyzed according to the new methods described in this work:
as a consequence, three events -- at energies 1968, 2061 and 2064~keV -- that were previously included in the analysis
window in Refs.~\cite{Agostini:2017iyd, Agostini:2018tnm, Agostini:2019hzm} are now discarded. 

The energy distribution of the events in the analysis window is fitted to search for
a signal due to $0\nu\beta\beta$ decay. The fit model includes
a Gaussian distribution for the signal, centered at $Q_{\beta\beta}$ with a width corresponding to the energy resolution, and a flat distribution for the background.
The free parameters of the fit are the signal strength $S=1/T_{1/2}$ and the
background index $B$.
The expectation value of the number of signal events scales with $S$ as
\begin{equation}
\mu_s = \frac{\ln{2} \, \mathcal{N}_A}{m_{76}} \, \varepsilon \, \mathcal{E} \, S \mathrm{,}
\label{mus} 
\end{equation}
where $\mathcal{N}_A$ is Avogadro's number, $m_{76}$ the molar mass of $^{76}$Ge, $\mathcal{E}$ the exposure and
$\varepsilon$ the total efficiency of detecting $0\nu\beta\beta$ decays.
The average $0\nu\beta\beta$ decay detection efficiency of each detector type and its breakdown in individual components
are listed in Tab.~\ref{tab:datasets}.
The mean number of background events in the analysis window is given by
\begin{equation}
\mu_b = B \times \Delta E \times \mathcal{E} \mathrm{,}
\label{mub}
\end{equation}
with $\Delta E = 240$~keV being the net width of the analysis window.
Data of each detector are divided in partitions, i.e. periods of time in which parameters are stable. Each partition $k$
is characterized by its own energy resolution $\sigma_k = \mathrm{FWHM}/2.35$, efficiency $\varepsilon_k$ and
exposure $\mathcal{E}_k$ . The signal strength $S$ and the background index $B$ instead are
common parameters to all partitions. This construction is a significant improvement compared to the analysis used in the
past~\cite{Agostini:2017iyd, Agostini:2018tnm, Agostini:2019hzm} as it allows a precise tracing of the performance
of each detector at any given moment. Furthermore, the background index is now
assumed to be the same for all detectors, while independent parameters for each detector type were used previously.
This change is motivated by the lack of any statistically significant indication
of a different background depending on detector type, position within the array, or time.

The statistical analysis is based on an unbinned extended likelihood function and it is performed
in both frequentist and Bayesian frameworks, following the procedure described in~\cite{Agostini:2017iyd}.
The likelihood function is given by the product of likelihoods of each partition, weighted with
the Poisson term:
\begin{eqnarray}
  \mathcal{L} &=& \prod_k \left[ \frac{(\mu_{s,k}+\mu_{b,k})^{N_k}\,e^{-(\mu_{s,k}+\mu_{b,k})}}{N_k!} \times \right. \\
  & & \left. \prod_{i=1}^{N_k}   \frac{1}{\mu_{s,k} + \mu_{b,k}} \times \left(\frac{\mu_{b,k}}{\Delta E} + \frac{\mu_{s,k}}{\sqrt{2 \pi} \sigma_k}e^{-\frac{(E_i-Q_{\beta\beta})^2}{2 \sigma_k^2}}\right) \right] \nonumber
\end{eqnarray}
where $E_i$ is the energy of the $N_k$ events in the $k$-th partition. The parameters $\mu_{s,k}$ and $\mu_{b,k}$ are calculated from
Eqs.~(\ref{mus}) and (\ref{mub}) and are partition-dependent.  
Phase~I data sets are included in the analysis as individual partitions with independent background indices.

The frequentist analysis is performed using a two-sided test statistics based on the profile likelihood.
The probability distributions of the test statistic are computed using Monte Carlo techniques, as
they are found to significantly deviate from $\chi^2$ distributions.
The analysis of the $N=13$ events of Phase~II yields no indication for a signal and a lower limit of
$T_{1/2} > 1.5 \times 10^{26}$~yr at 90\% C.L. is set.
Phase~I and Phase~II data together give a total exposure of 127.2~kg\,yr, which corresponds to $(1.288 \pm 0.018)$~kmol\,yr
of $^{76}$Ge in the active volume. The combined analysis has also a best fit for null signal strength, and provides a half-life limit of
\begin{equation}
T_{1/2} > 1.8 \times 10^{26}\textrm{~yr at 90\% C.L.}
\end{equation}
The limit coincides with the sensitivity, defined as the median expectation under the
no signal hypothesis.

GERDA achieved an unprecedentedly low background in Phase~II, as derived from the fit, of
$B=5.2^{+1.6}_{-1.3} \times 10^{-4}$~counts/(keV\,kg\,yr), and met the design goal
of background-free performance: 
the mean background expected in the signal region $(Q_{\beta \beta} \pm 2 \sigma)$ is 0.3 counts.

The statistical analysis is carried out also within a Bayesian framework.
The one-dimensional posterior probability density function $P(S|data)$ of the
signal strength is derived by marginalizing over the other free parameters by 
using the Bayesian analysis toolkit BAT~\cite{Schulz:2020ebm}.
The prior distribution for $S$ is assumed to be constant between 0 and
$10^{-24}$~1/yr, as in previous GERDA works. 
The limit on the half-life from Phase~I and II together is $T_{1/2} > 1.4 \times 10^{26}$~yr (90\%~C.I.).
A stronger limit $2.3 \times 10^{26}$~yr (90\%~C.I.) is obtained assuming a priori equiprobable Majorana
neutrino masses $m_{\beta\beta}$ (as $S \propto m_{\beta \beta}^2$), instead of equiprobable signal strengths.

Uncertainties on the energy reconstruction, energy resolution, and efficiencies are folded into the analysis through additional nuisance parameters, each constrained by a Gaussian probability distribution. Their overall effect on the limit is at the percent level.
Potential systematic uncertainties related to the fit model are found to marginally impact the results.
For instance, the limit changes by a few percent if a linear energy distribution is assumed for the background.

\begin{figure}
    \includegraphics[width=0.98\columnwidth]{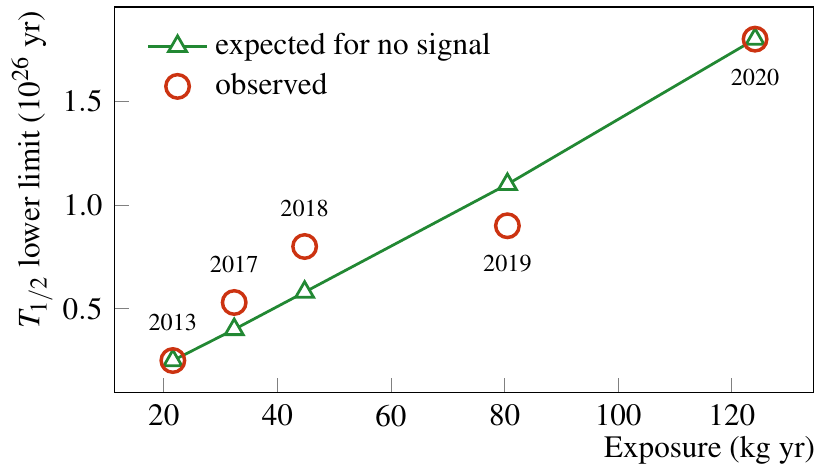}
    \caption{Circles: lower limit (90\% C.L.) on the $0\nu\beta\beta$ decay half-life of $^{76}$Ge set by GERDA as a function of the exposure~\cite{Agostini:2013mzu,Agostini:2017iyd,Agostini:2018tnm,Agostini:2019hzm}. Triangles: median expectation in the assumption of no signal.}
    \label{fig:history}
\end{figure}

Fig.~\ref{fig:history} shows the improvement achieved by GERDA with increasing exposure for the measured lower limit on the $0\nu\beta\beta$ decay
half-life of $^{76}$Ge and for the sensitivity. 
The background-free regime results in a linear improvement of sensitivity vs. exposure. GERDA is the experiment
providing the best sensitivity and the most stringent constraint on the half-life of any $0\nu\beta\beta$ decay.

The $T_{1/2}$ limit can be converted into an upper limit on the effective Majorana neutrino mass under the assumption that the decay is dominated by the exchange of light Majorana neutrinos.
Assuming a standard value of $g_A=1.27$, the phase space factor and the set of nuclear matrix 
elements from Refs.~\cite{Rodriguez:2010mn, Mustonen:2013zu, Vaquero:2014dna, Horoi:2015tkc, 
Hyvarinen:2015bda, Barea:2015kwa, Menendez:2017fdf, Song:2017ktj, Simkovic:2018hiq, Fang:2018tui, 
Coraggio:2020hwx}, a limit of $m_{\beta\beta} < 79-180$~meV at 90\% C.L. is obtained, 
which is comparable to the most stringent constraints from other
isotopes~\cite{Anton:2019wmi,KamLAND-Zen:2016pfg,Adams:2019jhp}.

GERDA has been a pioneering experiment in the search for $0\nu\beta\beta$ decay. GERDA improved the sensitivity by one order of
magnitude with respect to previous $^{76}$Ge experiments~\cite{KlapdorKleingrothaus:2000sn,Aalseth:2000ud} 
and proved that a background-free experiment based on $^{76}$Ge is feasible. Indeed, the 
LEGEND Collaboration~\cite{Abgrall:2017syy} is preparing a next generation experiment with a
sensitivity to the half-life of $0\nu\beta\beta$ decay up to $10^{28}$~yr.
In the first phase, LEGEND-200 has taken over the GERDA infrastructure at LNGS and will start data taking in 2021.


The data shown in Fig.~\ref{fig:spectrum} and the data relevant for the GERDA
Phase~II statistical analysis are available in ASCII format as
Supplemental Material \cite{Note1}.

\begin{acknowledgments}
The GERDA experiment is supported financially by
the German Federal Ministry for Education and Research (BMBF),
the German Research Foundation (DFG),
the Italian Istituto Nazionale di Fisica Nucleare (INFN),
the Max Planck Society (MPG),
the Polish National Science Centre (NCN),
the Foundation for Polish Science (TEAM/2016-2/17),
the Russian Foundation for Basic Research,
and the Swiss National Science Foundation (SNF).
This project has received funding/support from the European Union's
Horizon 2020 research and innovation programme under
the Marie Sklodowska-Curie grant agreements No 690575 and No 674896.
The institutions acknowledge also internal financial support.
The GERDA Collaboration thanks the directors and the staff of
the LNGS for their continuous strong support of the GERDA experiment.
\end{acknowledgments}


\bibliographystyle{apsrev4-1}

\addcontentsline{toc}{chapter}{Bibliography}

\end{document}